\documentclass[12pt,preprint]{aastex} 

\usepackage{url}

\newcommand{\kmps}{\hbox{~\rm{$\mbox{km s}^{-1}$}}}

\newcommand{\asr}{ {\it Ann. Rev. Astron. Astrophys.}}
\newcommand{\jaa}{ {\it J. Astrophys. Astron.}}

\shorttitle{Transient Magnetic and Doppler Velocity Features}
\shortauthors{Maurya, Vemareddy and Ambastha}

\begin{document}

\title{Velocity and Magnetic Transients Driven by the X2.2 White-Light Flare of 2011 February 15 in NOAA 11158}

\author{R. A. Maurya$^{1,2}$, P. Vemareddy$^1$ and A. Ambastha$^1$}
\affil{$^1$Udaipur Solar Observatory, Physical Research Laboratory, Udaipur-313001, India.}
\affil{$^2$Astronomy Program, Department of Physics and Astronomy, Seoul National University, Seoul 151-747, Korea.}

\email{ramajor@astro.snu.ac.kr, vema@prl.res.in, ambastha@prl.res.in}


\begin{abstract}

The first X-class flare of the current solar cycle 24 occurred in
Active Region NOAA 11158 during its central meridian passage on 2011
February 15. This two ribbon white-light flare was well observed by
the Helioseismic and Magnetic Imager (HMI) on board Solar Dynamics
Observatory (SDO). During the peak phase of the flare, we detected
magnetic and Doppler velocity transients appearing near the umbral
boundary of the main sunspot. These transients persisted for a few
minutes and showed spatial and temporal correspondence with the
flare kernels. The observed magnetic polarity at the transients'
locations underwent sign reversal, together with large enhancement
in Doppler velocities. We explain this observational phenomena using
the HMI spectral data obtained before, during and after the flare.
These changes were reflected in the maps of the AR in all the Stokes
parameters. Association of the transient features with various
signatures of the flare and the cause and effects of their
appearance are also presented on the basis of present theoretical
models. 
\end{abstract}



\newpage
\section{Introduction}
\label{S-introd}

It is known that catastrophic changes in the magnetic field
configuration of active  regions (ARs) at coronal heights trigger
energetic explosive events such as flares and coronal mass ejections
(CMEs). Therefore, variations in magnetic and Doppler velocity
fields are expected to accompany these energetic events. It was
first suggested by \citet{Giovanelli1939} that energy release in
flares should be associated with observable magnetic field changes.
Extensive efforts have been made since then to detect such changes
using photospheric magnetic field observations. Several types of
changes in the measured magnetic fields from pre- and post-phases of
flares have been reported by ground-based \citep{Patterson1981,
Patterson1984, Wang1992, Ambastha1993, Chen1994, Hagyard1999}, and
more recently by space-based \citep{Kosovichev2001, Qiu2003,
Wang2006, Maurya2008, Maurya2009, Maurya2010} observations. These
changes in observed magnetic fields can be divided into two main
categories: ``permanent'' and ``transient''. The first type is
irreversible change observed during pre- to post- phases of flares.
The later, ``transient'' changes are reversible that occur only
during the peak or impulsive phase of energetic events.

The observed changes in magnetic field parameters during the
impulsive phase of a flare are, however, expected to be affected by
a variety of effects that could introduce ambiguity in the cause and
effects of the observed changes. One major concern is that the flare
associated modification of spectral line profiles, used for the
measurement of photospheric magnetic fields, may lead to incorrect
estimation of magnetic fields during the impulsive phase of major
flares. Therefore, adequate care must be taken in interpreting the
``observed'' changes in photospheric magnetic fields and Doppler
velocities. For instance, the irreversible changes observed after
flares may occur due to flux emergence/cancelation process and may
be considered as ``real'' changes.  However, there is ambiguity in
interpreting the transients, reversible changes observed during some
large flares as it has been difficult to ascertain whether these
changes are real or artifacts. These ``magnetic transients'' are
therefore termed also as ``magnetic anomalies''
\citep{Patterson1984, Kosovichev2001, Qiu2003, Maurya2009}.

Some numerical experiments have been carried out to confirm magnetic
field changes due to a transient change of the spectral line profile
from absorption to emission \citep{Machado1980, Vernazza1981,
Ding1989, Ding2002, Edelman2004}. Effect of line profile change on
magnetic field estimation was also reported by \citet{Abramenko2004}
using spectrographic observations of six different photospheric
absorption lines at the flare and quiet locations. They showed that
the core of spectral profiles at the flare location was shallower as
compared to that for the quiet Sun, and less steep in the wings.
They attributed the enhanced core emission to thermal heating of
photosphere by the flare, and the less steep slope near wings to the
inhomogeneity of the photospheric magnetic fields. This resulted in
an underestimation of the magnetic field measurements by 18-25$\%$
at the flare locations. \citet{Qiu2003} have reported that sign
reversals of magnetic polarity may also be related to non-thermal
processes. They found anomalous polarity reversals during a large
X5.6 flare of 2001 April 6 at the location of strong hard X-ray
(HXR) emission forming near cooler umbral/penumbral regions of the
sunspots. These HXR sources are produced by energetic electron beams
impinging at the sites of higher density and strong magnetic fields.

In a recent study \citep[][hereafter MA09]{Maurya2009}, we had
discovered moving transient magnetic and Doppler velocity features
during the peak-phases of two major X17/4B and X10/2B white light
 flares (WLF) of 2003 October 28 and 29, respectively, in NOAA 10486.
 HXR sources were also observed during these flares. For the flare of 2003
October 28, the HXR sources showed a separation of $\approx23$ Mm
matching well with the chromospheric H$\alpha$ flare ribbons. This
observation is consistent with the flare models that suggest
formation of HXR footpoint sources in the chromosphere or lower
corona depending on the energy of penetrating particles. The
magnetic and Doppler velocity transients, on the other hand, showed
a somewhat greater separation of $\approx29$\,Mm similar to the
photospheric WLF kernels. At the transients' sites, we detected
anomalous magnetic polarity sign reversals during the impulsive
phases of these flares. The transients were first detected at the
weaker field location of polarity reversal line (PIL), and not in
the cooler, stronger magnetic field sites of sunspot umbra/penumbra,
as found earlier by \cite{Qiu2003}. However, as the flares in NOAA
10486 progressed, the transients moved away from the PIL towards the
stronger fields. We inferred from the analysis of the X17/4B flare
of 2003 October 28 that the moving transients were better related to
the HXR sources, i.e., the non-thermal processes associated with
electron-beam injection, and not to the WLF kernels, i.e., thermal
heating.

There have been a very few cases reported so far on the transient
magnetic and Doppler velocity features driven by the flares. It
appears that these transients may reach detectable levels only in
some very energetic, impulsive flares. In this paper, we report the
detection of a similar event associated with the first X-class flare
of the current solar cycle 24. This two-ribbon WL flare occurred in
NOAA AR 11158 during its central meridian passage on 2011 February
15. It was well observed by the Helioseismic and Magnetic Imager
(HMI) and Atmospheric Imaging Assembly (AIA) on board Solar Dynamics
Observatory (SDO). From the movies of high resolution magnetic and
Doppler  velocity images, we detected the transients appearing first
near the umbral boundary of the main sunspot. It persisted only for
a few minutes during the peak phase of the flare. As found earlier
in MA09, the magnetic polarity went through a sign reversal at the
location of the transients. The Doppler velocity, however, did not
show such a sign reversal. Instead, there occurred a large magnitude
velocity enhancement at the transient's site.

Using multi-wavelength flare observation from SDO, we intend to
study the association of the observed transients with various
signatures of the X2.2 flare of 2011 February 15. The spectral data
available from SDO-HMI will also be used to investigate the changes
in Stokes profiles occurring during the peak phase of the flare.
Characteristics of these transients will be examined to explain
their cause and effects using the present theoretical models. The
paper is organized as follow: In Section~\ref{S-DataAna},  we
describe the observational data and methods of analysis. Results and
discussions are presented in Section~\ref{S-ResDisc}. Finally,
summary and conclusions are given in the Section~\ref{S-Summ}.

\section{Observational Data and Analysis}
\label{S-DataAna}

AR NOAA 11158 was extensively observed by several space borne
instruments, such as, HMI \citep{Schou2012} and AIA
\citep{Lemen2012} on board SDO, Solar Optical Telescope
\citep[SOT,][]{Tsuneta2008} on board Hinode, and Reuven Ramaty High
Energy Solar Spectroscopic Imager \citep[RHESSI,][]{Lin2002}. We
have used the SOT images to study the flare evolution in the
chromosphere along with the spectral, magnetic field and Doppler
velocity data obtained from the SDO/HMI.

The HMI observes full solar disk in the photospheric absorption line
Fe\,{\sc i}  centered at the wavelength 6173.3\AA~and produces
images with spatial and temporal resolutions of
0\arcsec.5\,pixel$^{\rm -1}$ and 45\,s, respectively. HMI
magnetograms and Dopplergrams have precisions of 10\,G and 13
m\,s$^{-1}$, respectively. For examining the temporal changes in the
Fe {\sc i} line profile during the period of the flare, we have used
the 12 HMI intensity image data taken at six wavelength positions
from the line center, $\pm$34.4, $\pm$103.2 and $\pm$172.0\,m\AA, in
the two circular polarizations. Further, we have obtained the Stokes
profiles using the HMI Stokes parameters ({\it I}, {\it Q}, {\it U}
and {\it V}) at the aforesaid six wavelength positions. The spatial
resolution of these parameters is the same as the HMI magnetograms
while the temporal cadence is 215\,s.

The AIA observes the full Sun in several wavelengths  corresponding
to various layers of the solar atmosphere: 1700\AA~(continuum formed
in the photosphere and the temperature minimum region at the
temperature, log\,$T=$ 3.7); 304\AA~({He\,\sc ii} formed in the
chromosphere and the transition region at the temperature, log\,$T=$
4.7); 193\AA~({Fe\,\sc xx, xxiv} formed in the coronal and hot flare
plasmas at the temperature, log\,$T=$ 6.2, 7.3). The spatial and
temporal resolutions of these observations are 0\arcsec.6
pixel$^{-1}$ and 45\,s, respectively.

We have used the Ca{\sc ii} H 3968\AA~line images obtained from the
SOT to study the flare kernels at the chromospheric level. The
Ca{\sc ii} H line data have spatial and temporal resolutions of
0\arcsec.1 pixel$^{\rm -1}$ and 20\,s, respectively. These images
were scaled, registered and co-aligned with the corresponding HMI
magnetograms.

We extracted 180$\times$180\,pixel$^2$ areas-of-interest, centered
around the AR NOAA 11158 (Carrington longitude 36\arcdeg, latitude
-21\arcdeg), from the full disk HMI white-light intensitygrams,
magnetograms, dopplergrams, and the AIA images. We also extracted
the corresponding line intensity profiles and Stokes vectors. To get
images aligned within sub-pixels accuracy, we have used Fourier
transform based registration method implemented in the Interactive
Data Language (IDL).

The Fourier modulated HXR data during the flare  were obtained from
the RHESSI data archive \citep{Lin2002}. RHESSI observes the Sun in
the energy range 3\,keV to 17\,MeV, with a spatial resolution of
$\sim$2\arcsec.3 in the full Sun field-of-view. We have used the
``clean'' algorithm implemented under the Solar SoftWare (SSW), to
reconstruct the HXR images. The basic method was first developed for
radio astronomy by \citet{Hogbom1974}. It is an iterative algorithm
based on the assumption that the image can be well represented by a
superposition of point sources. Details of this method to
reconstruct the images from the RHESSI data are described in
\citet{Hurford2002}.

For studying the association of the transients driven by the X2.2
flare with HXR sources, we have constructed the RHESSI HXR images in
the energy band 25-50\,keV with spatial and temporal resolutions of
4\arcsec\,pixel$^{-1}$ and 45\,s, respectively, keeping the image
center fixed at the position (205\arcsec, -222\arcsec). Association
of HXR sources with the Ca{\sc ii} H flare ribbons and the
transients are discussed in the Section~\ref{Sb-HXR}.

\section{Results and Discussions}
\label{S-ResDisc}

\begin{figure}
\centering
\includegraphics[width=1.0\textwidth, clip=,bb=35 15 526 352]{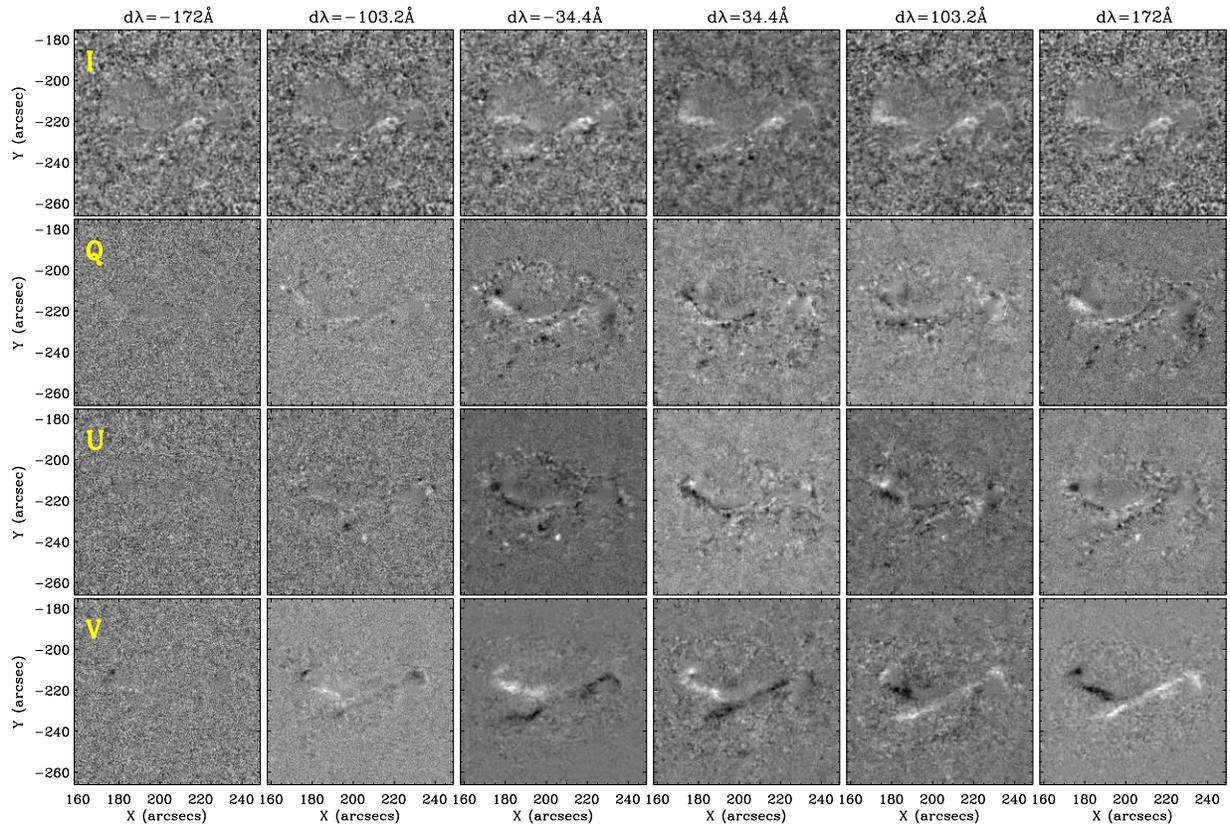}
\caption{Difference maps of the AR NOAA 11158 corresponding to the
Stokes parameters I, Q, U and V (from top to bottom rows) at the six
HMI wavelength positions (from left to right) at the peak phase
(01:52:30\,UT) of the flare from the pre-flare phase
(01:45:45\,UT).} \label{StkDiffMap}
\end{figure}

Figures~\ref{TimSeqImg} -- \ref{StkProf} show the results obtained
from the  analysis of the observational data for the AR NOAA 11158
during the X2.2 flare of 2011 February 15. In the following, we
discuss these results in detail.

\subsection{Morphology of AR NOAA 11158 and the X2.2 Flare}
\label{S-AR1158}

AR NOAA 11158 appeared near the disk center at location S19W03 on
2011 February 12 in the rising phase of the  current solar cycle 24.
Subsequent to its birth on February 12, it evolved very rapidly and
quickly developed from a simple $\beta$- to a very complex
$\beta\gamma\delta$-configuration by 2011 February 15. (A detailed
study of the surface and sub-surface evolution of this AR has been
undertaken and results therefrom will be communicated in a
subsequent paper.)

During its disk transit, NOAA 11158 produced several C-class and
5\,M-class flares.  It also unleashed the first X-class flare of the
current solar cycle observed on 2011 February 15. This event was a
two ribbon WLF as seen in HMI white-light images. The flare peaked
at 01:56\,UT, when the AR was centered around S20W12, and was
registered X2.2 magnitude on the Richter scale of solar flares. In
Figure~\ref{TimSeqImg}(top two rows), development of the flare is
shown using SOT Ca{\sc ii} H filtergrams of the AR NOAA 11158 on
2011 February 15.

The co-aligned, extracted HMI and AIA images of AR NOAA 11158,
obtained in different wavelengths during the flare, are shown in
Figure~\ref{ImMoz}. The top panel shows the AIA images taken at the
rising phase of the flare around 01:46 UT in 1700, 304 and
193\AA~(from the left to right). These wavelengths correspond to
different layers of the solar atmosphere, viz., temperature minimum
region, chromosphere and corona, respectively. The bottom panel
shows the HMI white-light image, magnetogram and dopplergram taken
around the peak phase of the flare at 01:53 UT.

Animations of HMI magnetograms and dopplergrams showed ``magnetic''
and faint ``Doppler'' transient features (TFs) appearing near the
umbral boundary of the main sunspot during the peak phase of the
X2.2 flare. The dotted (dashed) curves labeled by L1 and L2
represent the locations of the TFs during the peak phase of the
flare, at 01:53\,UT,  when the features were prominently seen in the
HMI magnetogram. These curves are drawn over the subsequent figures
for reference. It is evident from the bottom panel that these
transients appeared in rather narrow ridges along the umbral
boundaries of the opposite polarity sunspots. The images in the top
panel further illustrate that L1-L2 were associated with the flare
ribbons observed in different AIA wavelengths.

\begin{figure}
\centering
\includegraphics[width=1.0\textwidth, clip=,bb=11 8 447 353]{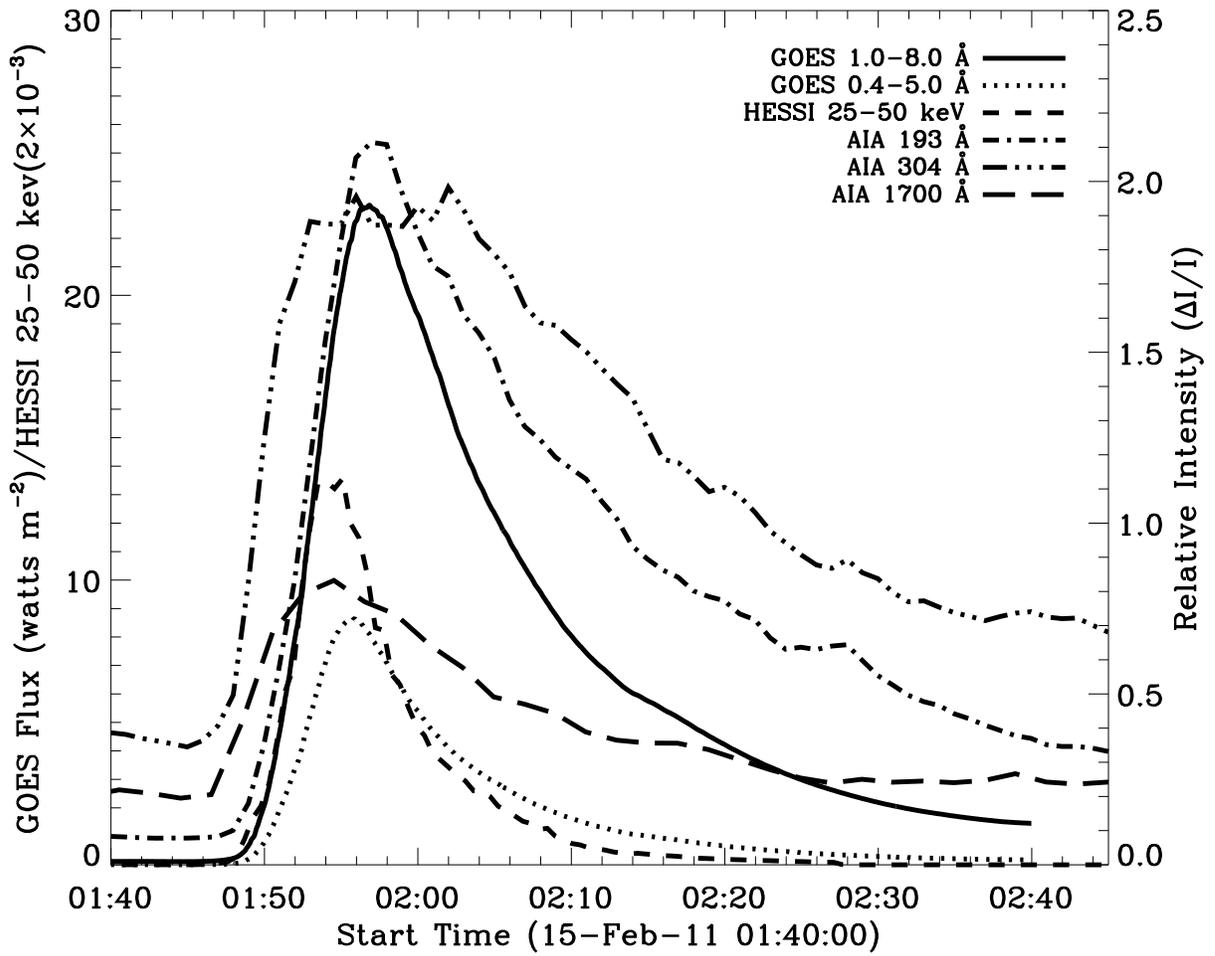}
\caption{Intensity profiles of the X2.2 flare of 2011 February 15 in
different wavelengths.} \label{LCurvs}
\end{figure}

Profiles of relative mean flare intensity $\Delta I/I$ with respect
to the quiet Sun in different AIA wavelengths, obtained from the
extracted and co-aligned images, are shown in Figure~\ref{LCurvs}.
GOES soft X-ray (SXR) profiles in the wavelength range 1.0-8.0
(0.4-5.0\AA) are shown by solid (dotted) curves. RHESSI HXR light
curve (dashed) in the energy band 25-50\,keV is also shown, which is
scaled in the plot range as marked in the left-hand side ordinate.
GOES 1.0-8.0\AA~and GOES 0.4-5.0\AA~emissions peaked at 01:56:49 and
01:55:50\,UT, respectively, while the RHESSI HXR emission peaked a
few minutes before, at 01:53:40.

The relative flare intensity profiles obtained in different AIA
wavelengths are also plotted in Figure~\ref{LCurvs}, according to
the scale given in the right-hand side ordinate. As evident, the
flare intensities in the AIA wavelengths of 193, 304 and
1700\AA~started rising around 01:46:53, 01:44:41 and 01:45:44\,UT,
and peaked at 01:58, 01:56, 01:54 UT, respectively. The flare
decayed, in all the wavelengths, by 02:40 UT, i.e., after one hour
of the onset of the flare. It is to note that the AIA emissions,
particularly in 304\AA, saturated during the peak phase of the
flare. Hence it is difficult to precisely identify the flare maximum
in these wavelengths. However, it is evident that the AIA intensity
profiles showed similar trend as the GOES SXR, while, the HXR
profile was much narrower with steep rise and decay phases.

The light curves in different wavelengths follow the standard flare
model for energy propagation at different heights in the solar
atmosphere \citep{Kane1974}. The general understanding of the flare
physics is based on the concept of reconnection of magnetic field
lines at coronal heights \citep{Sturrock1966, Hirayama1974,
Kopp1976}. Energy released during a flare covers a wide range of
wavelengths, however, it is easier to observe in certain spectral
lines as shown in Figure~\ref{LCurvs}. During the pre-flare stage,
where the release of the stored magnetic energy is triggered, the
chromospheric flare ribbons are seen easily (see,
Figure~\ref{TimSeqImg} {\it top}). The increasing separation of
flare ribbon with time results from reconnection of magnetic field
lines at successively increasing coronal heights. Further, the
energetic particles released in the corona during a flare take
increasingly longer time to propagate through the denser layers
downwards to the photosphere. This agrees with the observed flare
light curves in 304\AA~ which started rising at 01:44:41\,UT and
thereafter in 1700~\AA~corresponding to the photosphere and the
temperature minimum region. As the flare progressed, the temperature
also increased making the flare ribbons visible in higher
temperature plasmas in the coronal regions corresponding to the
193\AA~emission.

\subsection{The Magnetic and Doppler Velocity Transients}
\label{Sb-MagVel}

\begin{figure}
\centering
\includegraphics[width=1.0\textwidth,clip=,bb=51 32 517 479]{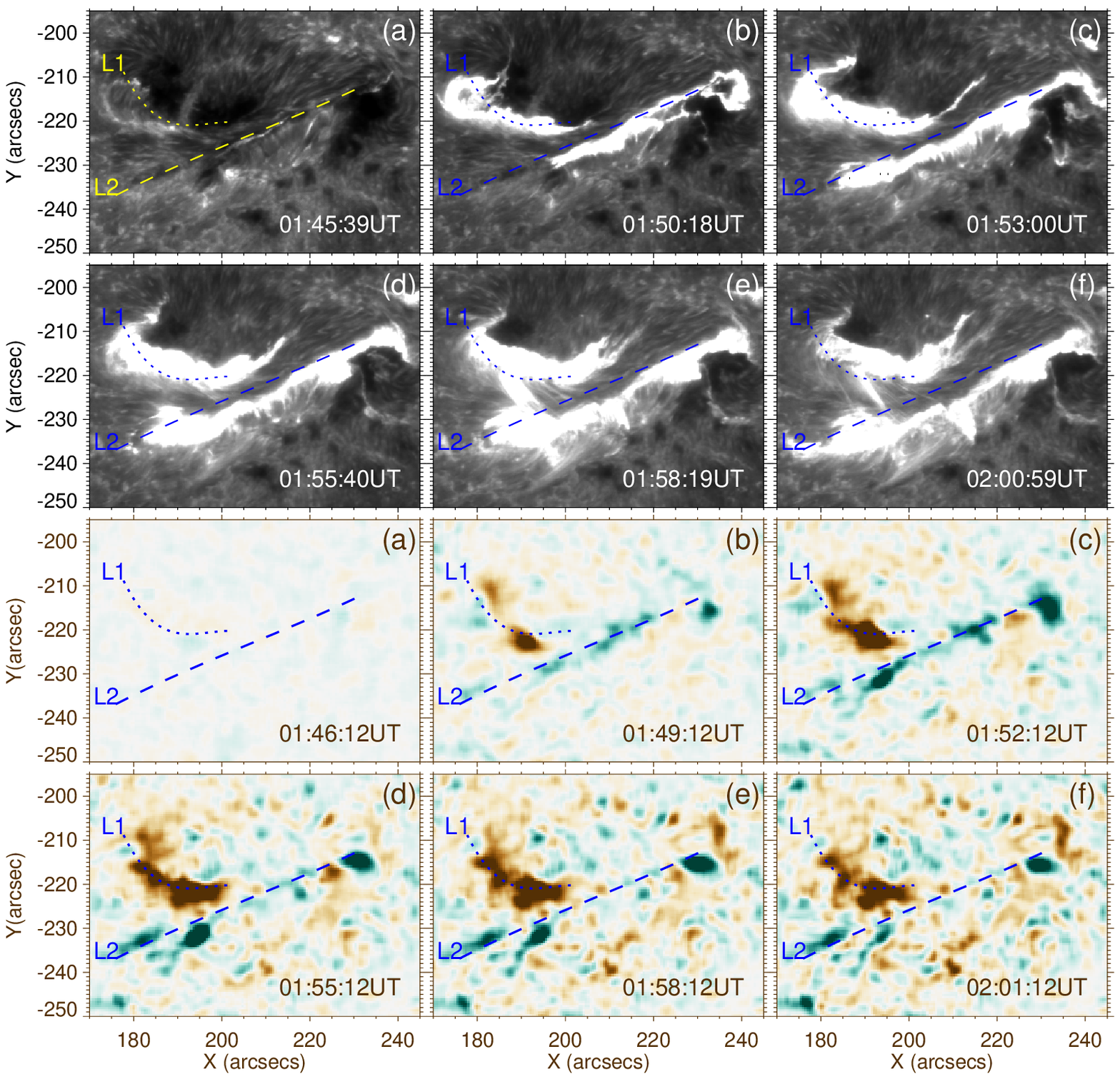}
\caption{(Top two rows): Ca{\sc ii} H filtergrams showing the
spatio-temporal  development of the two-ribbon X2.2 flare in AR NOAA
11158 on 2011 February 15. (Bottom two rows): Consecutive difference
images of HMI magnetograms showing the enhanced ``magnetic
transient'' features associated with the flare during the same
period as above. Dotted and dashed curves marked by L1 and L2
represent the locations of the transients, drawn for reference,
during the impulsive phase of the flare at 01:53 UT .}
\label{TimSeqImg}
\end{figure}

\begin{figure}
\centering
\includegraphics[width=1.0\textwidth,clip=,bb=50 30 515 333]{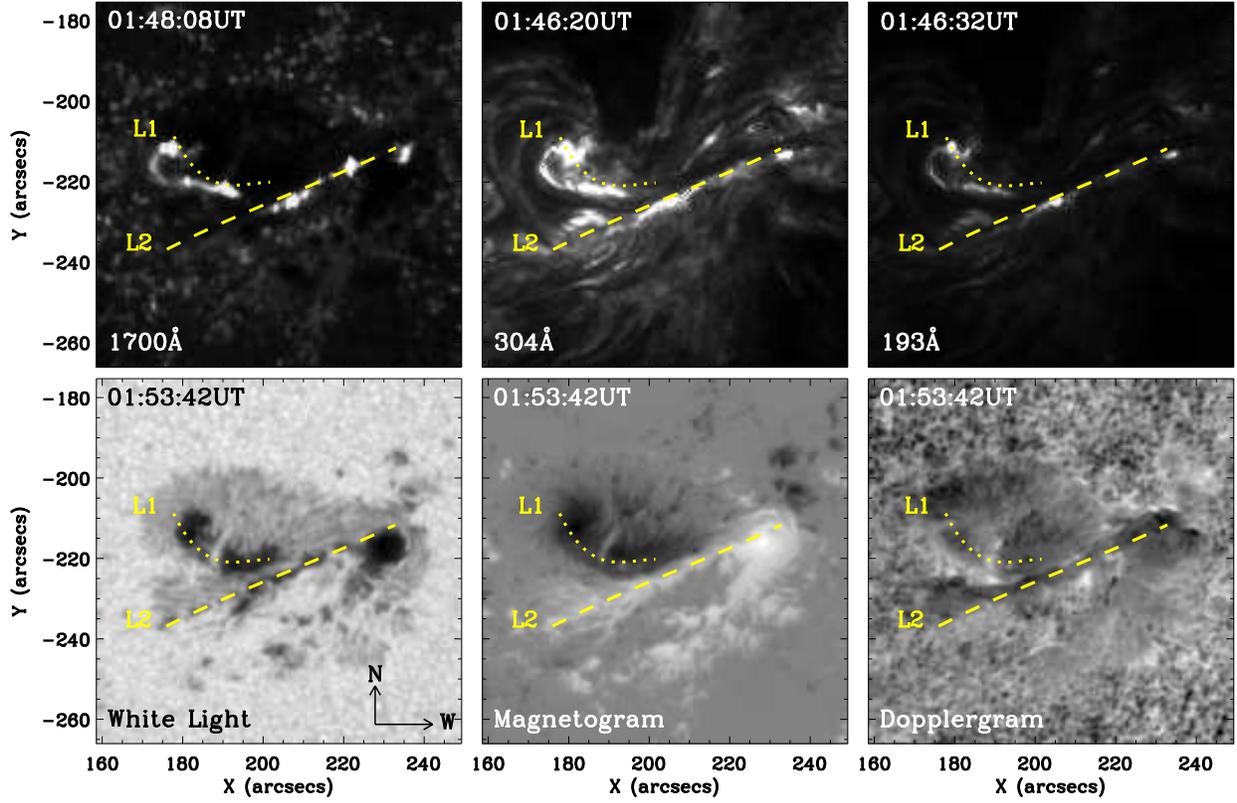}
\caption{AR NOAA 11158 during the X2.2 flare of 2011 February 15.
\textit{Top (from left to right):} AIA images in the wavelengths
1700, 304 and 193\AA. \textit{Bottom (from left to right):} HMI
white-light, magnetogram and dopplergram images. Dotted and dashed
curves marked by L1 and L2 represent the locations of the
transients, drawn for reference, during the impulsive phase of the
flare at 01:53 UT.} \label{ImMoz}
\end{figure}

The observed ``magnetic'' and ``Doppler'' TFs appeared during the
peak phase of the X2.2 flare of 2011 February 15 (c.f.,
Figure~\ref{ImMoz}). Similar features were reported earlier during
the more energetic X17/4B and X10/2B flares of October 28 and 29,
2003 \citep{Maurya2008, Maurya2009,Maurya2010d}, and the X5.6 flare
of 2001 April 6 \citep{Qiu2003}. \citet{Venkatakrishnan2008} have
also reported co-spatial Doppler ribbons associated with the
H$\alpha$ flare ribbons of the X17/4B flare. These TFs, usually
located around cooler umbral boundary of the sunspots, were found to
be largely co-spatial with the flare ribbons observed at different
heights of the solar atmosphere, viz, chromosphere, transition
region and corona (see Figures~\ref{TimSeqImg},~\ref{ImMoz}).

A time sequence of the consecutive difference images of the HMI
magnetograms of AR NOAA 11158 in Figure~\ref{TimSeqImg} shows dark
patches representing the magnetic transients along the curves L1 and
L2. These TFs appeared and faded in a few minutes' period
(01:50-02:02\,UT) during the impulsive phase of the flare. From
Figure~\ref{TimSeqImg}, it is evident that the transients were
co-spatial with the observed Ca{\sc ii} H line flare ribbons.
However, while the Ca{\sc ii} H flare ribbons separated away, the
TFs remained nearly stationary with time.

Using an automated method described in \citet{Maurya2010b}, we
determined that the Ca{\sc ii} H flare ribbons separated out with an
average velocity of 8 \kmps. This is much smaller as compared to the
earlier reported velocities in the range of 50-75 \kmps~ for the
flare ribbons and the TFs observed during the X17/4B super-flare of
28 October 2003 in NOAA 10486. In that event, the TFs rapidly moved
away from the weak field regions of the PIL reaching a maximum
separation of around 27.9$\pm$0.4\,Mm. Another part of the TF
observed near the leading sunspot (i.e., strong magnetic field
region), on the other hand, remained stationary (MA09). Similarly,
the TFs associated with the X10/2B flare of October 29, 2003 in the
same AR occurred in the umbral/penumbral region of the following
sunspot. Those TFs also remained stationary as observed here in the
case of the X2.2 flare of 2011 February 15.

\subsubsection{Anomalous Reversal of Magnetic Polarity}
\label{SSb-PolRev}

\begin{figure}
\centering
\includegraphics[width=1.0\textwidth,clip=,bb=13 12 528 480]{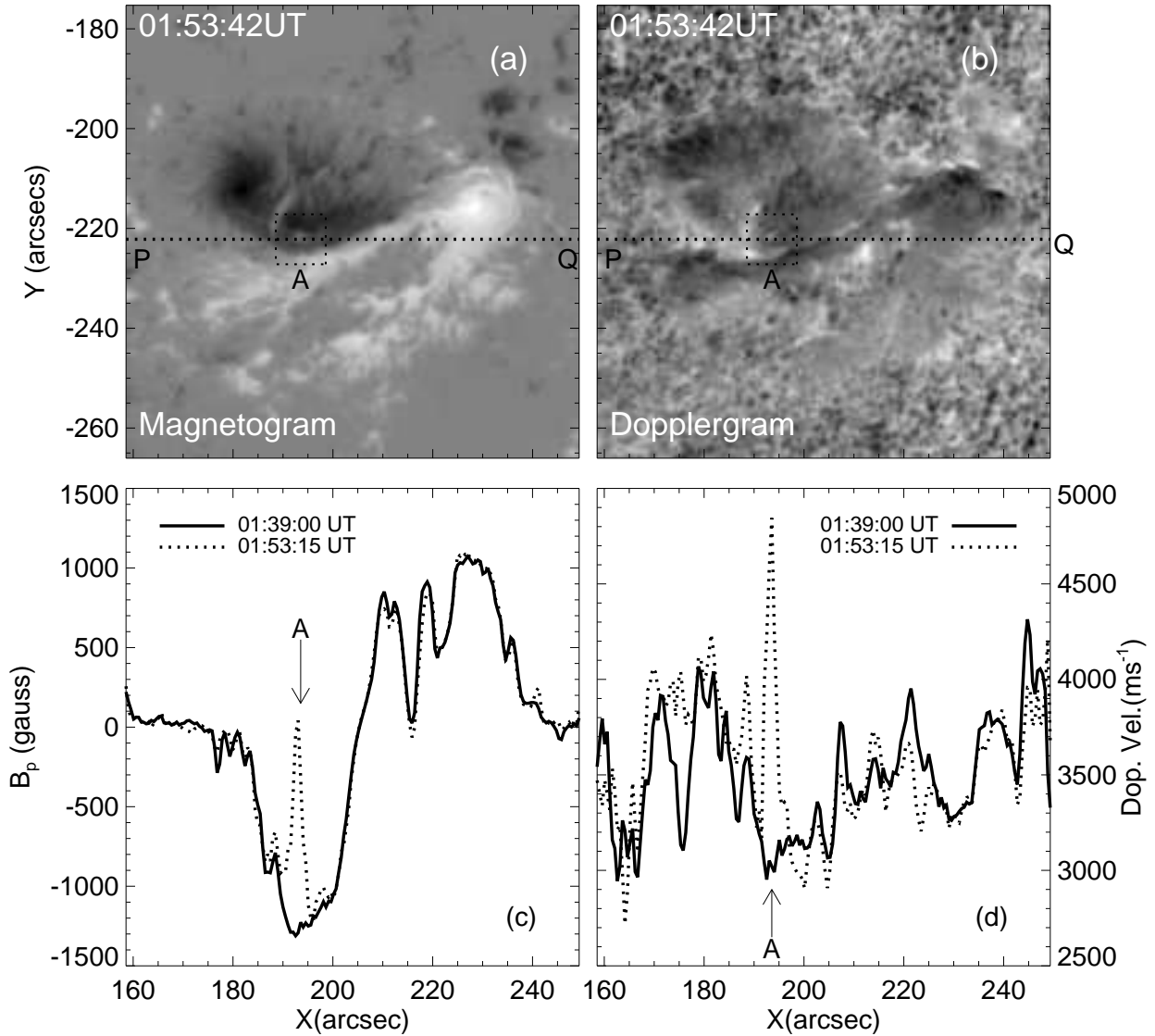}
\caption{AR NOAA 11158 at the peak phase of the X2.2 flare of 2011
February 15: (a) the magnetogram, and (b) the dopplergram. The box
labeled by ``A'' marks the area where TFs were observed. The bottom
panels (c) and (d) show the corresponding profiles along the line PQ
at the pre- and peak-phases of the flare represented by solid and
dotted curves, respectively.} \label{MDProf}
\end{figure}

We measured magnetic flux and Doppler velocity in AR NOAA 11158
along a horizontal line PQ (Figure~\ref{MDProf},\textit{top}) drawn
through the location `A' of the observed TF before and during the
flare maximum in order to investigate their variations. The line PQ
was selected by plotting the magnetic flux profiles along a
horizontal raster moving from the bottom to the top of the selected
magnetograms. The profiles of magnetic flux and Doppler velocity
along PQ are shown in Figure~\ref{MDProf} (\textit{bottom}), where
the solid and dashed curves represent the magnetic fluxes/velocities
before (01:39:00\,UT) and during (01:53:15\,UT) the flare maximum,
respectively. The magnetic flux profiles during the pre- and
peak-phases of the flare matched at all points along PQ except at
the location `A' (see Figure~\ref{MDProf}c). It gives a clear
evidence of an abnormal sign reversal in magnetic flux polarity at
`A' around the peak phase of the flare from the pre-flare phase. The
Doppler velocity profiles along PQ obtained during the pre-  and
peak phase of the flare, however, do not match so well as is the
case for the magnetic flux profiles (see Figure~\ref{MDProf}d). This
is mainly due to the effect of solar p-mode oscillations. But, there
is also a large increase in the Doppler velocity during the peak
phase of the flare at `A', i.e., the location of sign reversal in
magnetic polarity.

What processes caused the observed abnormal magnetic flux sign
reversal and the Doppler velocity enhancement? Are they real
changes? Or, is it related to line profile change \citep{Ding2002,
Qiu2003} or some other process such as the impact of a shock wave
\citep{Zharkova2007} on the Fe {\sc i} line-production region? More
recently, \citet{Kosovichev2011a} have reported ``sunquake'' sources
in the regions of TFs and suggested these to arise due to thermal
and hydrodynamic effects of high-energy particles heating the lower
atmosphere. Apart from the observed transients in magnetic flux and
Doppler velocities, increase in the continuum intensity was seen in
the form of white-light flare ribbons. Notably, the flares studied
in MA09 were also white light events during which similar TFs were
detected.

One may attribute the observed sign reversals to the expected change
in the spectral line profile from absorption to emission occurring
during these large WL flares. However, the non-LTE calculations by
\citet{Ding2002} were carried out for the spectral line Ni {\sc i}
6768\AA, used in GONG and MDI instruments for magnetic and Doppler
measurements. They showed that this line, formed in thermally stable
regions, can turn into emission only by a large increase of electron
density and not by heating of the atmosphere by any other means.
They suggested that such non-thermal effects are most pronounced in
a cool atmosphere where continuum is maintained at low intensity
level. The observed flare-associated sign reversal in magnetic flux
reported here also occurred in the cooler umbral boundary of the
sunspot in agreement with \citet{Ding2002}. However, there are also
exceptions as reported in the case of the X17/4B flare of October
28, 2003 in NOAA 10486, where the TFs formed in the weak magnetic
field regions around the PIL.

If the line formation model given by \citet{Ding2002} for Ni {\sc i}
were to be applicable also to the \mbox{Fe {\sc i}} line used in the
HMI measurement, one would suggest that the observed sign reversal
is due to the change in line profile. An observational evidence of
the Ni {\sc i} line profile change was not available for the WL
flares previously studied in MA09 due to the lack of spectral data
from MDI and GONG. Therefore, our interpretation of the origin of
the transients was limited to their spatio-temporal association with
WL flare kernels and non-thermal HXR sources. However, as the
required spectral data is now available from the HMI for the X2.2
flare studied here, we can look for the observational evidence of
any flare related changes in the Fe {\sc i} line profiles at the
location of the TFs. The results are discussed in
Section~\ref{Sb-LineProf}.

\subsubsection{Space-Time Maps of Magnetic flux and Doppler Velocity}
\label{SSb-STD}

In order to study the temporal variations of the TFs, we constructed
space-time maps of magnetic flux and Doppler velocity in the AR
along the line PQ as shown in Figure~\ref{MDProf}(\textit{top}). We
added corresponding magnetic and Doppler velocity values along PQ
during the period 01:38:15--02:15:45\,UT of 2011 February 15. We
also constructed a similar space-time map for the HMI white-light
images.

\begin{figure}
\centering
\includegraphics[width=1.0\textwidth,clip=,bb=32 1 533 360]{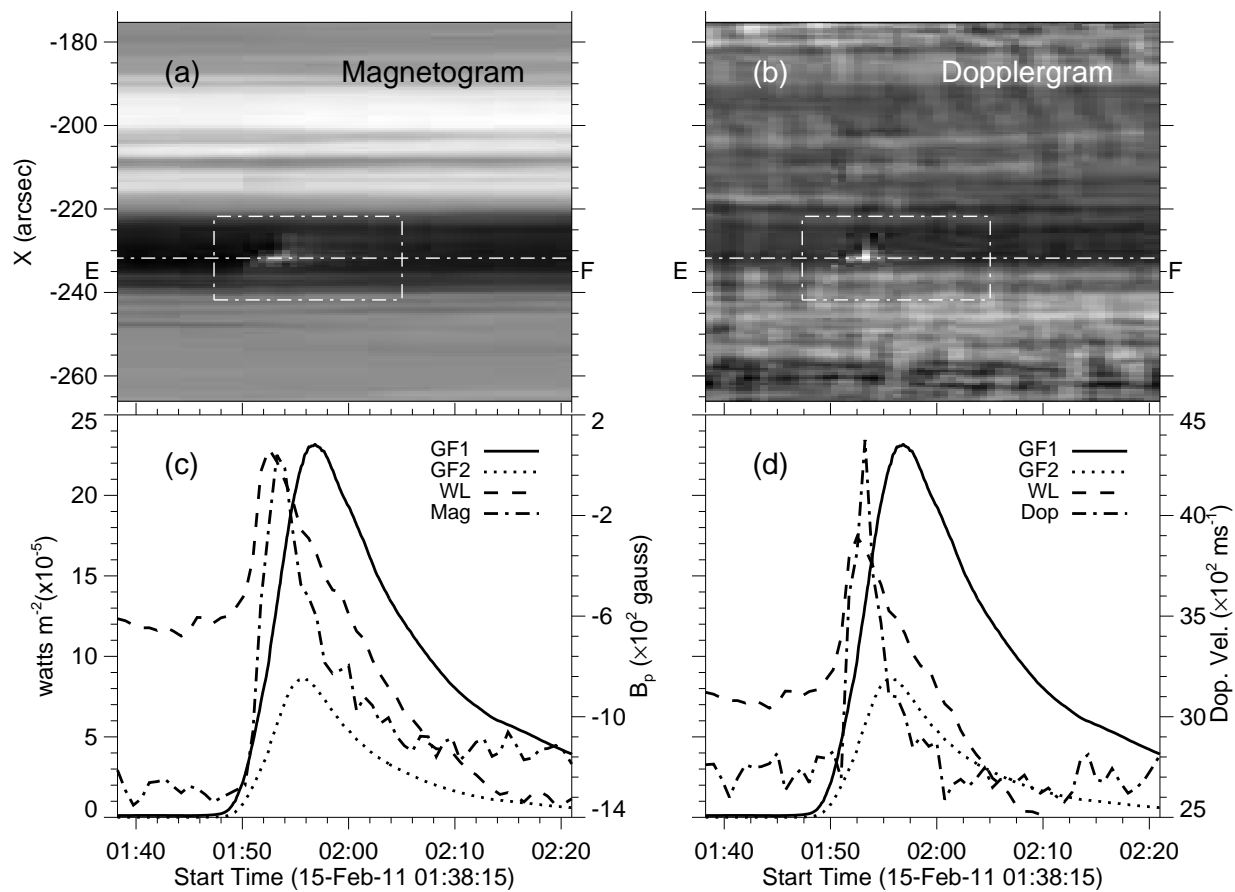}
\caption{Space-time maps of (a) magnetic flux, and (b) Doppler
velocity along the line PQ of Figure~\ref{MDProf}(a-b). The boxes
enclose the ranges of time and location of the TFs seen during the
flare. Corresponding bottom panels show the dashed-dotted profiles
along the line EF for (c) magnetic flux, and (d) Doppler velocity.
Also plotted are the WLF intensity (dashed), integrated GOES flux
GF1(GF2) in the wavelength range 1.0--8.0(0.5--4.0)\AA~ represented
by solid (dotted) curves.} \label{STMap}
\end{figure}

The space-time maps thus constructed are shown in
Figure~\ref{STMap}(\textit{top}). Here, the x- and y-axes represent
time and longitudinal positions along PQ, respectively. As observed
during the period 1:47-2:08\,UT, these maps show a clear signature
of the magnetic and Doppler transients in the box drawn along the
time line EF at the longitude 195\arcsec. Magnetic flux (MF) and
Doppler velocity (DV) values along the line EF  are plotted as
dashed-dotted curves in the corresponding bottom panels. The solid
(dotted) curves show the integrated GOES-15 SXR flux in the
wavelength range 1.0--8.0 (0.5--4.0) \AA, while the dashed curves
represent the WLF intensity scaled in the left-hand side y-axis.

It is evident from Figure~\ref{STMap}(\textit{bottom}) that MF and
WLF began rising at 01:47\,UT with a delay of one (two) minutes from
the rise of the GOES flux in the wavelength range
1.0-8.0(0.4-5.0)\AA. However, MF reached its peak much faster, at
01:53:15\,UT, as compared to various other flare intensities: WLF at
01:54:00\,UT, and GOES SXR 0.4--5.0 (1.0--8.0)\AA~at 01:55:50
(01:56:49)\,UT. Similarly, the MF and WLF profiles decayed much
faster than the GOES flux. The DV profile also peaked simultaneously
with the WLF and MF profiles. Its profile returned back to the
original level at 2:00\,UT, i.e., about 10 minutes before the
magnetic flux. From these time profiles at `A' (Figure~\ref{MDProf},
\textit{top}), we thus infer that the observed transients and WLF
had similar temporal profiles, reaching their peaks 2-3 min prior to
the peaks for the GOES fluxes. Also, the DV profile was
comparatively much sharper than the MF profile.

If the MF and DV transients occurred due to line profile changes
\citep{Qiu2003}, then they both are expected to exhibit similar
temporal evolution. Similarity of the transient and WLF profiles
suggests that thermal heating process may have caused these
transients. The time delay between the TFs and GOES SXR peaks
indicates the travel time that the energetic charged particles
released in corona during the flare took to reach at the photosphere
and produce the TFs and subsequent thermal heating giving rise to
the soft X-ray flux. We further discuss the spatial and temporal
variations of TFs along the curves L1 and L2, and their association
with WLF, Ca{\sc ii} H and HXR kernels in the following section.

\subsection{The White-light, Ca{\sc ii} H Flare and the HXR Sources}
\label{Sb-HXR}

\begin{figure}
\centering
\includegraphics[width=1.0\textwidth,clip=,bb=60 30 467 517]{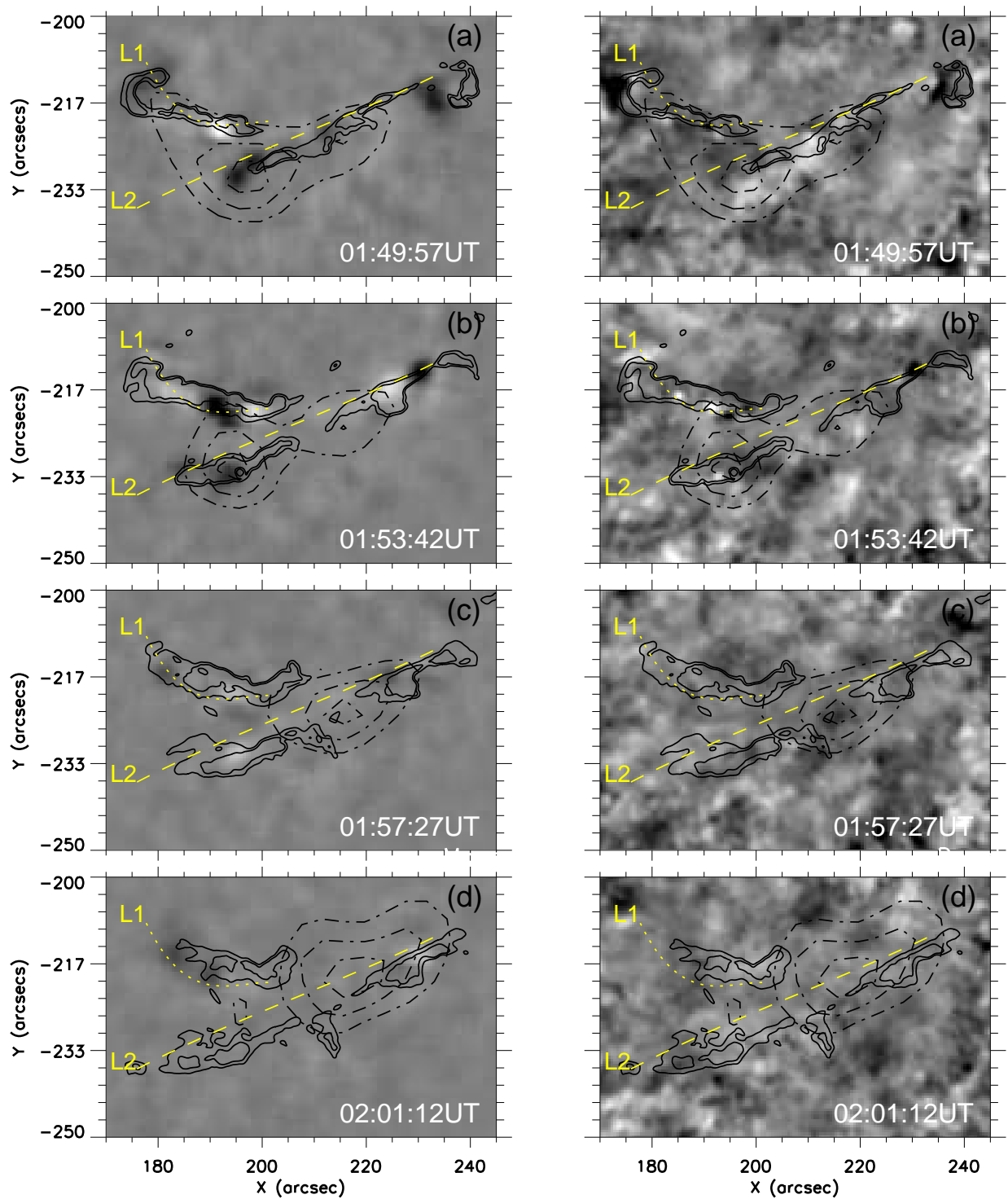}
\caption{Temporal sequence of consecutive difference images: (left
column) HMI magnetogram,   and (right column) Dopplergram with
overlaid RHESSI HXR (dashed dotted) and Ca{\sc ii} H flare intensity
(solid) contours. L1 and L2 represent the reference locations of the
TFs at 01:53 UT.} \label{DiffMap}
\end{figure}

In order to identify the temporal and spatial association of RHESSI
HXR  sources and flare ribbons with the TFs, we overlaid the
25-50\,keV RHESSI HXR contours on the consecutive magnetogram and
dopplergram difference images (Figure~\ref{DiffMap}). The transients
are seen in the respective background images in the time period
1:49:57-02:01:12\,UT. The HXR contours (dash-dotted) were drawn at
40, 60 and 80\% levels of the maxima. These features were observed
around the magnetic polarity inversion line (PIL); not drawn here to
avoid overcrowding. The Ca{\sc ii} H flare ribbons (solid contours)
and the transients, i.e., L1 and L2 appear to be well correlated.

The HXR sources  evolved both spatially and temporally along L1 and
L2 as shown in the panels (a) to (d).  The source region of HXR
moved toward west, parallel to the magnetic neutral line during the
period when the TFs were seen. The observed motion of HXR sources is
attributed to the temporal movement of the reconnection site as
reported earlier by \citet{Masuda2001, Bogachev2005, Zhou2008}. From
Figure~\ref{DiffMap}, it is evident that the HXR sources were not
very well correlated with the TFs during the entire peak phase of
the flare as compared to the Ca{\sc ii} H ribbons. Therefore, we
infer that HXR sources alone were not the cause or effect of the
TFs. This inference agrees with our earlier results obtained for the
X10/2B flare of 2003 October 29 (MA09).

Although the WLF ribbons were faintly observable in the HMI
continuum animations, they are rather difficult to identify even in
the difference images. However, from the animations, the WLF ribbons
appeared to be correlated with the TFs. This is further corroborated
by the the WLF intensity, magnetic and Doppler velocity profiles of
Figure~\ref{STMap}(\textit{bottom}).

\citet{Qiu2003} had reported that (i) the apparent sign reversal
occurs  in cool, strong-field ($>$1000\,G) regions within sunspot
umbrae, (ii) locations of the anomaly are exactly co-aligned with
the thick-target HXR sources, and (iii) the transient reversal flux
is temporally correlated with the HXR flux. Based on these results,
they proposed that at the flare kernels the Ni {\sc i} absorption
line profile is either temporarily turned into emission or
significantly broadened with a strong central reversal due to the
non-thermal beam impact on the umbral atmosphere. However, our
results for this X2.2 flare are not in full conformity with
\citet{Qiu2003} as we did not find a very good association of HXR
and TFs. Therefore, the TFs observed during the flare of 2011
February 15, and the X10/2B flare of 2003 October 29 reported in
MA09, are not fully explained only by the non-thermal process
suggested in \citet{Qiu2003}.

\subsection{Line Profile Changes Associated with the Transients}
\label{Sb-LineProf}

Earlier studies based on numerical models \citep{Machado1980,
Vernazza1981,  Ding1989, Ding2002, Qiu2003} have shown that the
flare associated transients may be attributed to a change in the
spectral line profile. However, due to the non-availability of
observational spectral data, the transients could be examined using
only the imaged data of the flares at various wavelengths and
inferences were drawn based on this indirect method. We are now able
to address this issue using SDO/HMI spectral observations.  The HMI
observes the Sun in two circular polarizations at six wavelength
positions ($\pm34.4, \pm103.2$ and $\pm172.0$\,m\AA) of the {Fe \sc
i} spectral line. Using these observations one can construct the
line profile for any desired location of the full disk Sun. The
spectroscopic imaging of HMI observations are described in a recent
paper by \citet{MartinezOliveros2011}.

\begin{figure}
\centering
\includegraphics[width=1.0\textwidth, clip=,bb=38 30 445 397]{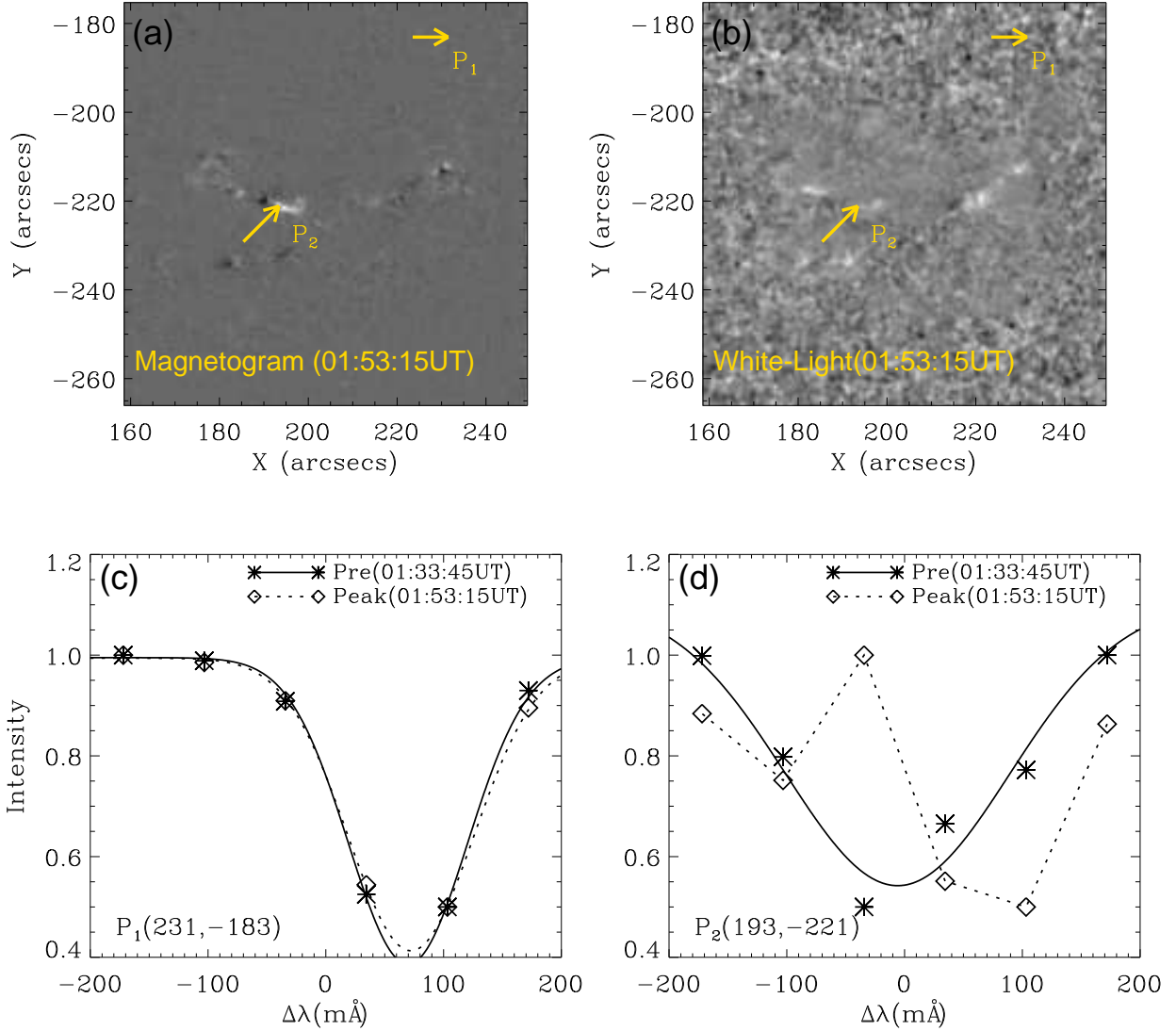}
\caption{Difference images of AR NOAA 11158 - (a) magnetogram, and
(b)Doppler velocity obtained at the peak phase (01:53:15\,UT) of the
flare with the pre-flare phase (01:33:45\,UT). LCP profiles of the
{Fe \sc i} line in (c) P$_1$ (quiet), and (d) P$_2$ (transient)
correspond to the locations marked by arrows in the top panel. The
profiles are shown for the pre (01:33:45\,UT) and peak
(01:53:15\,UT) phases, respectively by solid and dotted curves,
obtained from a four parameter Gaussian fitting of the six
wavelength positions, except for the peak phase at P$_2$.}
\label{LineProf}
\end{figure}

\begin{figure}
\centering
\includegraphics[width=0.9\textwidth, clip=,bb=25 18 368 514]{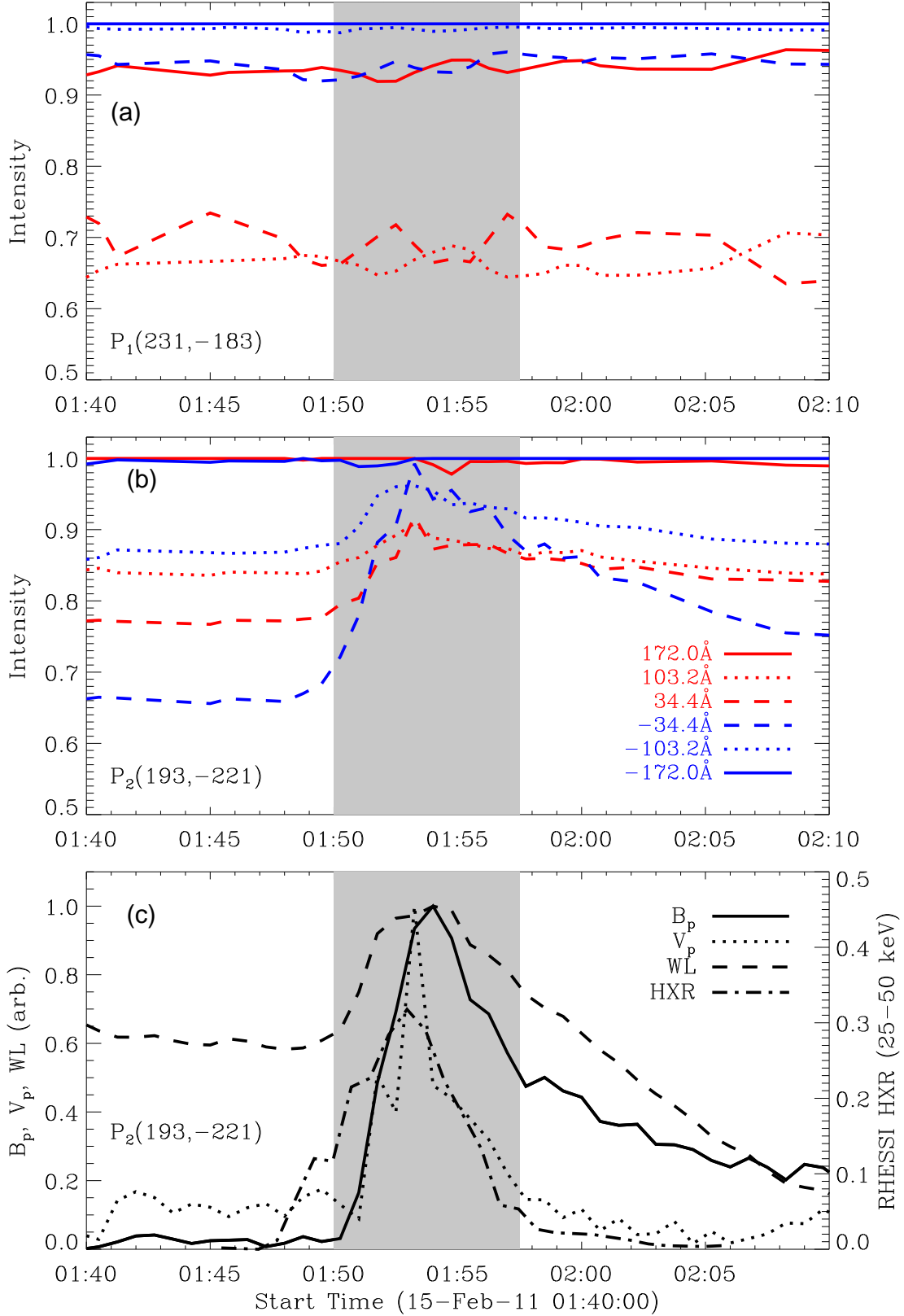}
\caption{Temporal variation of normalized LCP intensities at the six
wavelength positions of the Fe {\sc i} line for the locations (a)
P$_1$ and (b) P$_2$ (see Figure~\ref{LineProf}a, b). The
corresponding variations in the photospheric magnetic flux ($B_{\rm
p}$), Doppler velocity V$_p$, white light intensity (WL) and RHESSI
HXR energy at the location P$_2$ are shown in (c).}
\label{LineInten}
\end{figure}

\begin{figure}
\centering
\includegraphics[width=1.0\textwidth, clip=,bb=43 16 476 410]{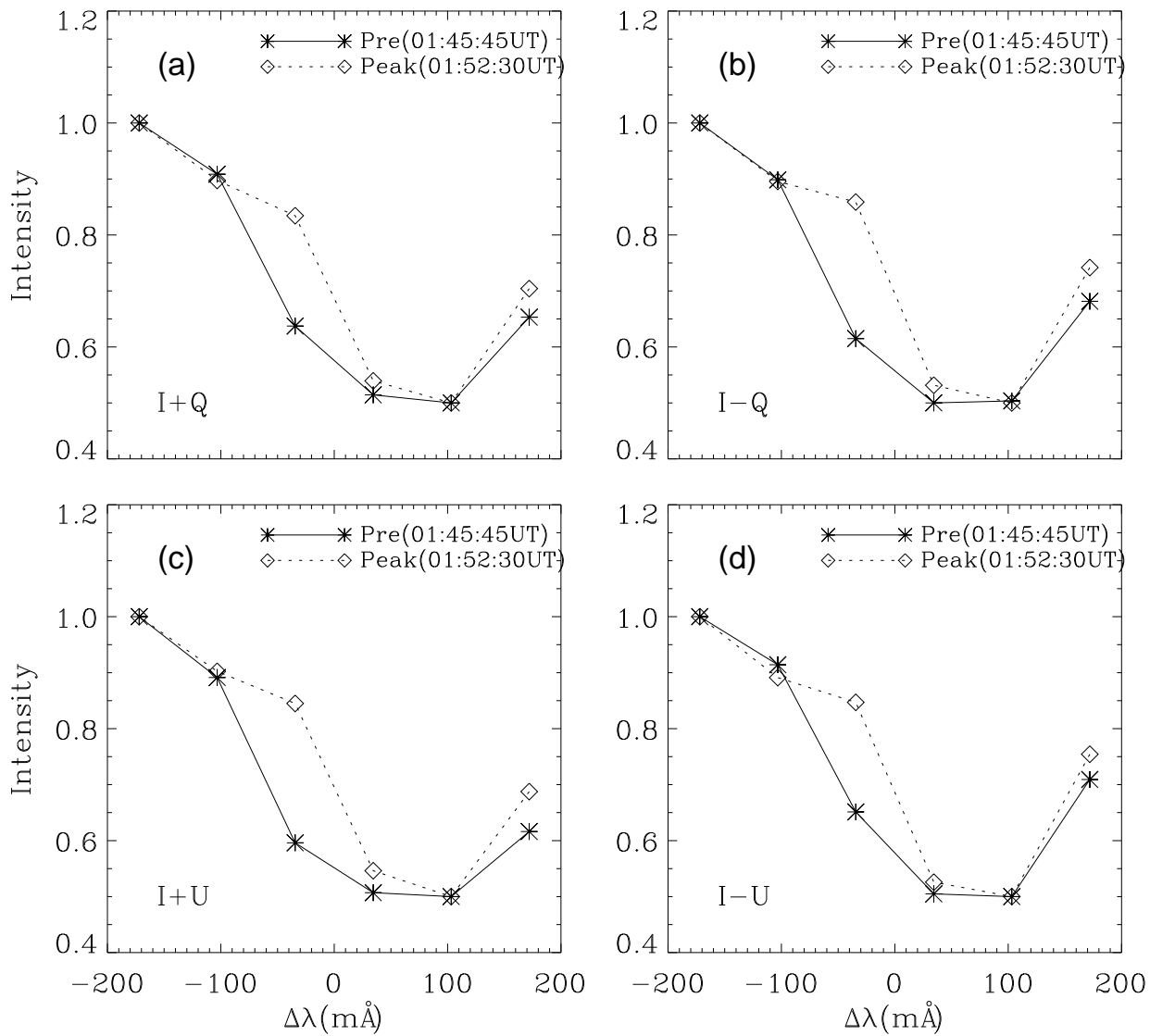}
\caption{Stokes profiles at the location P2 (see
Figure\ref{LineProf}(a)),  where solid (dotted) curves represent the
spectral profiles during the pre (peak) phase of the flare.}
\label{StkProf}
\end{figure}

We have already shown the strong spatial association of magnetic
transients with the WLF kernels. Figure~\ref{LineProf}(\textit{top})
shows the consecutive difference magnetogram  and white-light images
of the AR NOAA 11158 during the peak phase (01:53:15\,UT) of the
X2.2 flare. We selected a quiet (P$_1$) and a transient (P$_2$)
location as marked by the arrows. Figure~\ref{LineProf}(c-d) shows
the line profiles at these two locations. Solid (dashed) curve
represents the line profile during the pre (peak) phase of the
flare. These curves were obtained from a four term Gaussian fitting
(except for the peak profile at P$_2$ where such a fitting was not
possible). It is evident that the line profile turned to emission at
the wavelength $\Delta\lambda$=-34.4\,m\AA~and became broader during
the peak phase of the flare as compared to the pre-flare phase. This
change in the line profile shape is expected to be the main reason
for the observed flare associated sign reversal of magnetic flux and
the enhancement of Doppler velocity at the location P$_2$ during the
peak phase of the flare. No such change was found at P$_1$ taken as
a reference location far away from the flare.

From the HMI spectral line-profile data, we now attempt to explain
the reason for these transients. In the SDO/HMI measurements,
magnetogram and Dopplergram maps of the full disk Sun are computed
from the phase of the Fourier transform (first component) of the
line profile values. The phase is determined for the LCP and RCP
components independently, and the Dopplergrams and magnetograms are
constructed from the mean and difference, respectively
\citep{Schou2012}. However, the standard procedure assumes only a
moderate line shift due to Doppler velocity and Zeeman splitting.
The regions having large changes in the line profile shape, e.g.,
line reversal, and large shifts would not be covered by the
algorithm, resulting in artifacts in these measurements. For this
large X2.2 flare, the spectral line core at the locations of the
transient evidently turned from absorption to emission. Therefore,
it is difficult to get ``correct'' values of magnetic flux and
Doppler velocity using the above algorithm.

In the following, we have used a two point algorithm to illustrate
the result of the observed spectral line reversal at the transient
locations. The magnetic flux and Doppler velocity values can be
computed from the line intensities in the LCP and RCP as follow: Let
$I_i^{\rm LCP,RCP}$, $i=0, \ldots, 5$ be the line intensities in the
two circular polarizations, LCP and RCP, at the six wavelength
positions, $\Delta\lambda_i=68.8\,i-172.0$, centered around the Fe
{\sc i} spectral line. Then the mean intensities in the blue and red
wings of the line can be written as,

\[
 I_b^{\rm LCP, RCP}=\frac{1}{3}\sum_{i=0}^2{I_i^{\rm LCP, RCP}}
\]

\noindent and

\[
I_r^{\rm LCP, RCP}=\frac{1}{3}\sum_{i=3}^5{I_i^{\rm LCP, RCP}},
\]

\noindent respectively. The parameters sensitive to magnetic flux
(say, $\psi_B$) and Doppler velocity (say, $\psi_V$) fields, can be
written as,

\[
\psi_B=\frac{1}{2}\left[(I_b^{\rm RCP}+I_r^{\rm LCP})-(I_b^{\rm LCP}+I_r^{\rm RCP})\right]
\]

\noindent and

\[
\psi_V=\frac{1}{2}\left[(I_b^{\rm LCP}+I_b^{\rm RCP})-(I_r^{\rm LCP}+I_r^{\rm RCP})\right],
\]

\noindent respectively.

The parameters $\psi_B$ and $\psi_V$ for any location of the AR NOAA
11158  can thus be computed using the 12 line intensity observations
at six wavelength positions in the two circular polarizations.
However, it is to note that as such these parameters do not
represent the magnetic and Doppler velocity fields. They are defined
here only to examine the observed magnetic polarity sign reversal
and Doppler velocity enhancement at the transients' locations (see
Figure~\ref{MDProf}c, d). Furthermore, this is used to illustrate
the observed artifacts in magnetic and Doppler velocity measurements
resulting from the standard procedure adopted in SDO/HMI.

The values of the parameters $\psi_B$ and $\psi_V$ at the location
P$_2$  during pre(peak) phase of the flare was found to be
-0.034(-0.008) and 0.036(0.054), respectively. It is evident from
these computed values that the parameter $\psi_B$ tends towards a
positive value during the peak phase of the flare as compared to
that in the pre-flare phase. This indicates the reason for the
observed sign reversal in the magnetic polarity at location P$_2$ or
at A (see Figure~\ref{MDProf}c). Similarly, the velocity parameter
$\psi_V$ increased during the peak phase of the flare, indicating
the large enhancement in the HMI Doppler velocity at P$_2$ or at A
(see Figure~\ref{MDProf}d).

Figure~\ref{LineInten} shows the temporal variation in the
normalized  line intensities at the six wavelength positions
$\pm34.4, \pm103.2$ and $\pm172.0$\,m\AA, where panels (a) and (b)
correspond to the locations P$_1$ and P$_2$, respectively. Panel (a)
shows that there was no significant change in the line profile
shapes at P$_1$ from the pre- to peak-phase of the flare. The bottom
panel (c) shows the corresponding variation in the RHESSI HXR flux
(in the energy range 25-50\,keV), white-light (WL), photospheric
magnetic flux (B$_p$) and Doppler velocity (V$_p$). The parameters
B$_p$, V$_p$ and WL are normalized to the plot range at the left
ordinate axis. The RHESSI HXR fluxes are shown with respect to the
right ordinate axis. The shaded area shows the time range of the
transients. It is evident from panel (b) that the change in line
profile at P$_2$ is temporally associated with the transients. This
is also the period of the peak phase of the flare as seen from panel
(c). The spectral line intensity near the core (i.e. at wavelength
$\pm$34.4\,m\AA) increased towards the continuum. The intensity
enhancement was found to be stronger at the blue wing as compared to
that in the red wing. This asymmetry resulted in the enhancement of
Doppler velocity in the same sign (Figure~\ref{MDProf}d) as
explained earlier.

From the spectral data available during this flare, it is thus
evident that line profile changes occurred during the impulsive
phase of the flare. The estimated magnetic flux and Doppler velocity
values as obtained by HMI algorithms were affected by these changes
during the impulsive phase of the flare, resulting in the observed
transient features in magneto(Doppler)grams. It is to note that a
purely line-of-sight magnetic field would result in a spectral line
with a missing pi-component. Also, the horizontal and vertical
gradients in magnetic field can cause asymmetric line profiles. The
way to resolve these ambiguities would be to not just look at the
LCP and RCP profiles, but all the stokes profiles, i.e., {\it
I}$\pm${\it U} and {\it I}$\pm${\it Q}. Such profiles are now
available from the SDO/HMI. We obtained the required data
corresponding to this flare and aligned the Stokes vectors in a
similar manner as was done for other data sets of this study.
Figure~\ref{StkDiffMap} shows the consecutive difference maps of the
Stokes parameters ({\it I}, {\it Q}, {\it U} and {\it V})
constructed for the AR NOAA 11158 at the six wavelengths during the
peak phase of the flare. Strong transient signals are seen in all
these Stokes maps, which are particularly more prominent around the
Fe {\sc i} line center (i.e., at -34.4 and 34.4\,m\AA).

We have further analyzed the line profiles at several flare and
quiet locations of the AR NOAA 11158 for examining the transient
changes in  the Stokes profiles. These profiles {\it I}$\pm${\it U}
and {\it I}$\pm${\it Q} at the transient location P2 are plotted in
Figure~\ref{StkProf}. Solid (dotted) curves correspond to the
profiles during pre(peak) phase of the flare. There is a strong
reversal at the wavelength -34.4\,m\AA~away from the line center,
similar to the profile derived from the continuum as shown in
Figure~\ref{LineProf}(d). Here it is to note that the magnitude of
Stokes {\it I} is much larger at all wavelengths compared to the
magnitude of Stokes {\it Q} and {\it U}. Hence, the differences seen
in the computed Stokes profiles I$\pm$Q and I$\pm$U are small.

The observed line profile changes could be related to the
thermodynamic change occurring during the flare peak time as the
height of line formation can drastically change due to the moderate
perturbation of temperature and density.  This can be attributed to
a large increase of electron density as shown by the non-LTE
calculations of \citet{Ding2002} for the Ni {\sc i} 6768 \AA line.
They suggested that the non-thermal excitation and ionization by the
penetrating electrons generate a higher electron density which
enhances the continuum opacity, thereby pushing the formation height
of the line upward. The precipitation of electrons and deposition of
energy in the chromosphere, enhanced radiation in the hydrogen
Paschen continuum gives rise to the line source function, leading to
an increase of the line core emission relative to the far wing and
continuum.

\section{Summary and Conclusions} \label{S-Summ}

We have analyzed the HMI magnetic flux and Doppler velocity images
of AR  NOAA 11158 obtained during the X2.2 flare of 2011 February
15. We detected transient magnetic and Doppler features during the
peak phase of this flare as reported earlier in a few large X-class
flares. This transient phenomenon is not well understood as there
are questions on their physical mechanism and their association with
observed anomalous sign reversal in magnetic polarity, Doppler
velocity enhancement, WLF kernels and HXR sources, etc. From the
observational data obtained from different instruments, we derived
the following important characteristics of these transients driven
during the peak phase of the X2.2 flare.

The TFs showed spatial and temporal association with the flare
ribbons observed  at different heights in the solar atmosphere. They
were well correlated also with the WLF kernels. These features
occurred near the cooler, umbral/penumbra boundary of the main
sunspot. It is to note that the TFs detected in AR NOAA 10486 (MA09)
during the X17/4B flare of 2003 October 28 were also associated with
WLF kernels, and moved with velocities in the range of 30 to 50
\kmps. The TFs observed during the energetic X2.2 flare of 2011
February 15, however, were observed to be nearly stationary,
although the corresponding Ca{\sc ii} H line flare ribbons separated
out with a mean velocity of 8\,\kmps~ as calculated by an automated
method \citep{Maurya2010b}.

The anomalous sign reversal in magnetic polarity during the X2.2
flare of 2011 February 15 was found similar in nature as observed in
the AR NOAA 10486 during the X17/4B and X10/2B flares of 2003
October 28 and 29, respectively. It should be noted that the GONG
and MDI data for the flares in NOAA 10486 were based on the Ni {\sc
I} line, while the HMI data for the X2.2 flare studied in this paper
are based on the Fe {\sc I} line. Thus, the transient features
appearing in the peak phase of major flares seem to be unrelated to
the line used in the measurements.

The spectral data available from SDO-HMI provide unambiguous
evidence of the changes occurring in the Fe {\sc I} line and the
corresponding Stokes profiles during the impulsive phase of the
flare. The estimated magnetic flux and Doppler velocity values as
obtained by HMI algorithms were affected by these changes, resulting
in the observed transient features in magneto(Doppler)grams.

This change in the line profile may arise due to both thermal
effects and non-thermal excitation and ionization by the penetrating
electron jets produced during the large flares. The precipitation of
electrons and deposition of energy in the chromosphere gives rise to
the line source function leading to an increase of the line core
emission relative to the far wing and continuum of the Ni {\sc I}
line, as suggested by \citet{Ding2002}.

The magnetic transients were observed at the locations of anomalous
sign reversal of magnetic polarity. Co-temporal enhancement in the
measured Doppler velocity also occurred there. \citet{Qiu2003}
suggested that the observed sign reversal is caused by changes in
the line profile related to the non-thermal effects. But the TFs
seen during the X2.2 flare did not show a good correlation with the
HXR sources, particularly in the post-peak phase of the flare.
Therefore, energetic injection of electrons alone do not appear to
have caused the observed changes in the spectral line in this flare.
Thermal effects also appear to have contributed to the observed
changes as inferred from the observed association of the TFs with
the flare ribbons.

We thus conclude that the  magnetic and Doppler transients observed
during the X2.2 flare of 2011 February 15 were essentially related
to the line profile changes. These changes were reflected in the
maps of the AR in all the Stokes parameters. However, they did not
correspond to a `real' change occurring in the photospheric magnetic
flux or Doppler velocity during the impulsive phase of this, and the
other such large, energetic flares. Both thermal and non-thermal
physical processes operating during the flare may have contributed
to the transient changes in spectral line profiles. Therefore, the
observed magnetic and velocity transients may be considered to be
the observational signatures of these physical processes occurring
during the peak phase of the flare.

\acknowledgments This work utilizes data from the Helioseismic and
Magnetic Imager (HMI) and  the Atmospheric Imaging Assembly (AIA) on
board Solar Dynamics Observatory (SDO). This work also utilizes the
X-ray data obtained by Ramaty High-Energy Solar Spectroscopic Imager
(RHESSI), and Ca{\sc ii} H data from the Solar Optical Telescope
(SOT) on board Hinode. We are grateful to S. Couvidat of Stanford
University for providing the spectral data and helpful discussions.
We thank anonymous referee and Professor Jongchul Chae for their valuable comments and suggestions that helped in improving the manuscript. One of the authors (RAM) acknowledges support by the National Research Foundation of Korea (2011-0028102) under which a part of this work was carried out.



\newpage

\end{document}